\documentclass[a4paper,11pt]{article}
\usepackage{pos}
\usepackage{bm,latexsym,amsmath,mathrsfs}
\usepackage{graphicx}
\usepackage{color}
\usepackage{times,setspace}

\title{Toward dense QCD in quantum computers}
\author{Arata~Yamamoto}
\affiliation{Department of Physics, The University of Tokyo, Tokyo 113-0033, Japan}

\abstract{
Lattice QCD at nonzero baryon density is a big challenge in hadron physics.
In this presentation, I discuss the quantum computation of lattice gauge theory at nonzero density.
I show some benchmark results of the Schwinger model obtained by the quantum adiabatic algorithm and the quantum variational algorithm.
}

\FullConference{%
 The 38th International Symposium on Lattice Field Theory, LATTICE2021
  26th-30th July, 2021
  Zoom/Gather@Massachusetts Institute of Technology
}

\begin{document}
\maketitle

\section{Introduction}

Lattice QCD at nonzero density is very difficult because of the sign problem.
Many theoretical methods were proposed, but it still seems difficult in classical computers.
The hardness of the sign problem depends on temperature and chemical potential.
While the sign problem is mild at high temperature and small chemical potential, the sign problem is severe at low temperature and large chemical potential.
Zero temperature is the most difficult.
In this sense, the simulation at zero temperature is the most important mission.
It is also important from a phenomenological viewpoint, i.e., neutron-star physics.
I would like to propose the possibility to use quantum devices as a solution to this problem.
In particular, let's focus on nonzero density and zero temperature.

This presentation is based on the original paper~\cite{Yamamoto:2021vxp}.
In this proceedings paper, I would like to skip the technical details and to overview the outline.
For the details, see the original paper.

\section{Algorithms}

In the conventional lattice QCD, the Lagrangian formalism is used.
The Euclidean path integral is nothing but the thermal partition function, so chemical potential is naturally introduced.
On the other hand, the Hamiltonian formalism is favored for quantum simulation.
In the Hamiltonian formalism, we can work with fixed particle numbers.
Instead of thermal average, we consider the ground state because the ground state is the most relevant for zero-temperature physics.
Let's consider a certain lattice theory with fermions.
The lattice Hamiltonian $H$ and the fermion number operator $Q$ can be defined.
They commute with each other, $[H,Q]=0$.
This means that the ground state is labeled by the fermion number, as
\begin{equation}
 Q|\Psi(q)\rangle = q|\Psi(q)\rangle
,
\end{equation}
where $Q$ is the operator and $q$ is its eigenvalue.
What to do in the simulation at nonzero density is to obtain the ground state $|\Psi(q)\rangle$ for each particle number $q$ and then to calculate physical observables as a function of $q$.

\begin{figure}[h]
\centering
\includegraphics[scale=0.4]{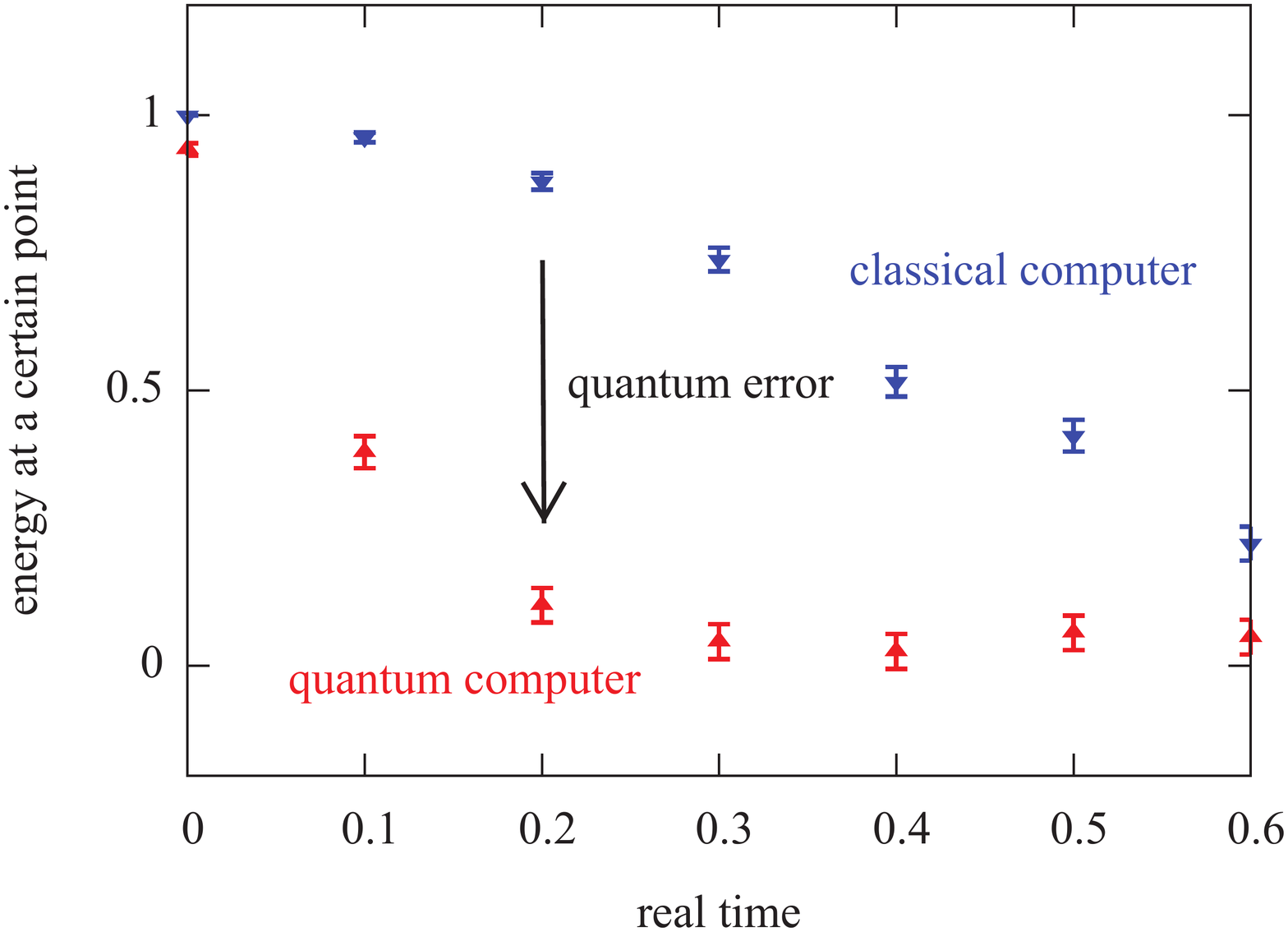}
\caption{\label{figqc}
The data obtained by a classical computer (blue) and the data obtained by a quantum computer (red).
The definitions of the vertical and horizontal axes are given in Ref.~\cite{Yamamoto:2020eqi}.
}
\end{figure}

How can we obtain the ground state in quantum computers?
The basic strategy is as follows.
First, we prepare the ground state $|\Psi_0(q)\rangle$ of a certain Hamiltonian $H_0$, which can be easily solved.
Then we manipulate quantum gates many times,
\begin{equation}
 |\Psi(q)\rangle = \prod_{s=1,\cdots,S}U(s) |\Psi_0(q)\rangle
,
\end{equation}
to give the ground state $|\Psi(q)\rangle$ of the full Hamiltonian $H$.
If all the evolution operators $U(s)$ commute with the fermion number operator $Q$, this process conserves the fermion number $q$.
There are two famous choices to construct the evolution operators: the quantum adiabatic algorithm and the quantum variational algorithm.
The quantum adiabatic algorithm is based on the adiabatic theorem in quantum mechanics \cite{Farhi}.
The operator is almost unique, and the convergence to the exact ground state is ensured in the limit of $S\to \infty$.
The quantum variational algorithm is a hybrid method of classical and quantum computers \cite{Peruzzo}.
The evolution operators contain several variational parameters.
The values of the parameters are tuned to minimize the total system energy.
In the quantum variational algorithm, $S$ can be small.
This final property is important for near-term projects of quantum computation.
The current quantum computer suffers from large quantum noise.
Figure \ref{figqc} is demonstration of the quantum noise.
Two kinds of data are plotted; one is obtained by a classical computer and the other is obtained by a quantum computer.
They are inconsistent due to uncontrollable quantum noises.
To suppress such artifact, the number of gate operations must be small.
For this reason, the quantum variational algorithm is favored in the current study of quantum computation.

\section{Benchmark tests}

I did the benchmark tests of these algorithms on the so-called ``simulator'', which is a classical computer to mimic a quantum computer.
I adopted the lattice Schwinger model on a small lattice $N=8$.
The Hamiltonian is
\begin{eqnarray}
 H &=& H_L + H_\chi
\\
 H_L &=& \sum_{n=1}^{N-1} L_n^2
\\
 H_\chi &=& -i \sum_{n=1}^{N-1} (\chi^\dagger_n U_n \chi_{n+1} - \chi^\dagger_{n+1} U^\dagger_n \chi_n)
\end{eqnarray}
and the fermion number operator is
\begin{equation}
 Q = \sum_{n=1}^{N} \left[ \chi_n^\dagger \chi_n - \frac12 \{1-(-1)^n\}\right].
\end{equation}
The initial Hamiltonian is taken as $H_0 = H_L$.
The evolution operators are set as
\begin{equation}
 U(s) = \exp \left[ -i\delta t \left(H_L+\frac{s}{S}H_\chi \right)\right]
\end{equation}
in the quantum adiabatic algorithm and 
\begin{equation}
 U(s) = \exp \left[ -i \left(\alpha H_L+\beta\frac{s}{S}H_\chi \right)\right]
\end{equation}
in the quantum variational algorithm.
Here $\delta t$ is a small step size and $\alpha$ and $\beta$ are the variational parameters.
Since the lattice size is small, the exact matrix diagonalization is possible.
I compared the simulation results with the exact answers.

\begin{figure}[h]
\begin{minipage}{0.5\linewidth}
\centering
\includegraphics[scale=0.75]{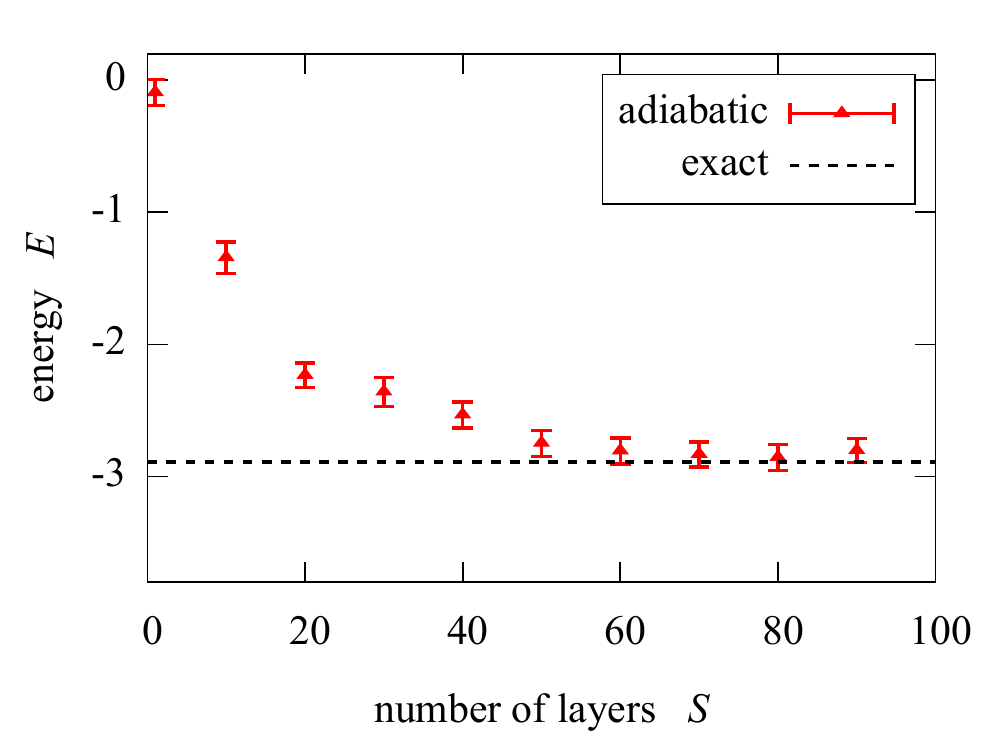}
\end{minipage}
\hfill
\begin{minipage}{0.5\linewidth}
\centering
\includegraphics[scale=0.75]{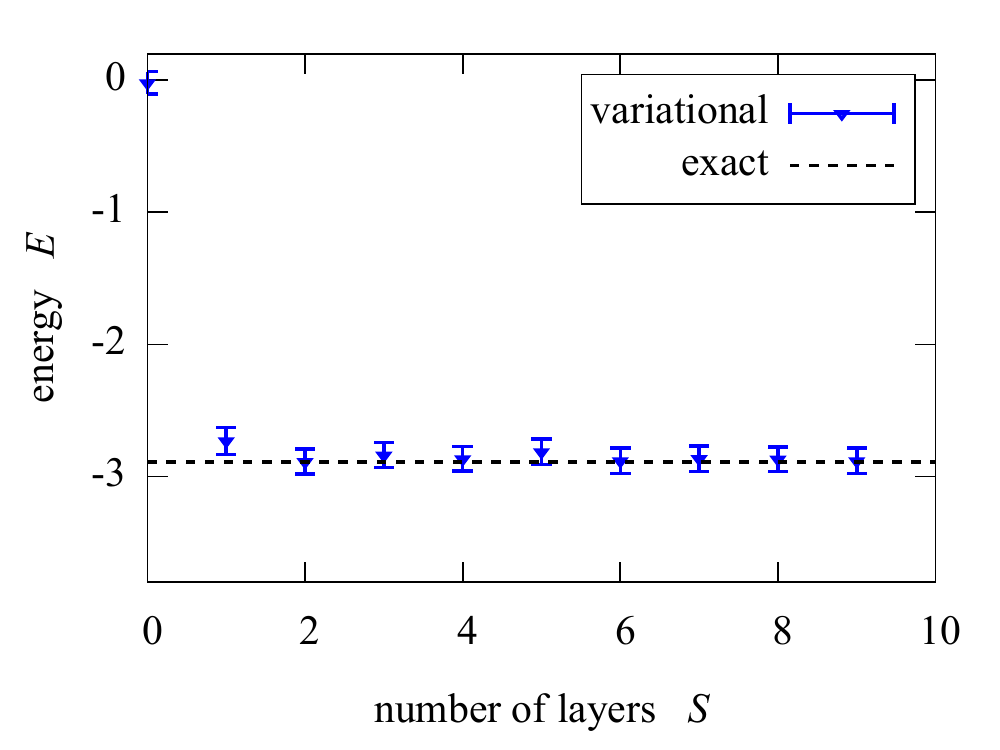}
\end{minipage}
\caption{\label{figE}
The total system energy $E$ calculated by the quantum adiabatic algorithm (left) and the quantum variational algorithm (right).
The dotted line is the exact value obtained by the matrix diagonalization.
The fermion number is $q=1$.
}
\end{figure}

In Fig.~\ref{figE}, the total system energy 
\begin{equation}
 E=\langle\Psi(q)| H |\Psi(q)\rangle
\end{equation}
is plotted as a function of $S$.
In both cases, the system energy is changed by the evolution operators, and eventually becomes flat.
Both of these algorithms successfully reproduce the exact value obtained by the matrix diagonalization.
The difference between them is the required value of $S$.
In the adiabatic method, the convergence is achieved around $S \simeq 60$.
In the variational method, the convergence is much faster; we need only one or two evolution operators.

\begin{figure}[h]
\centering
\includegraphics[scale=1.1]{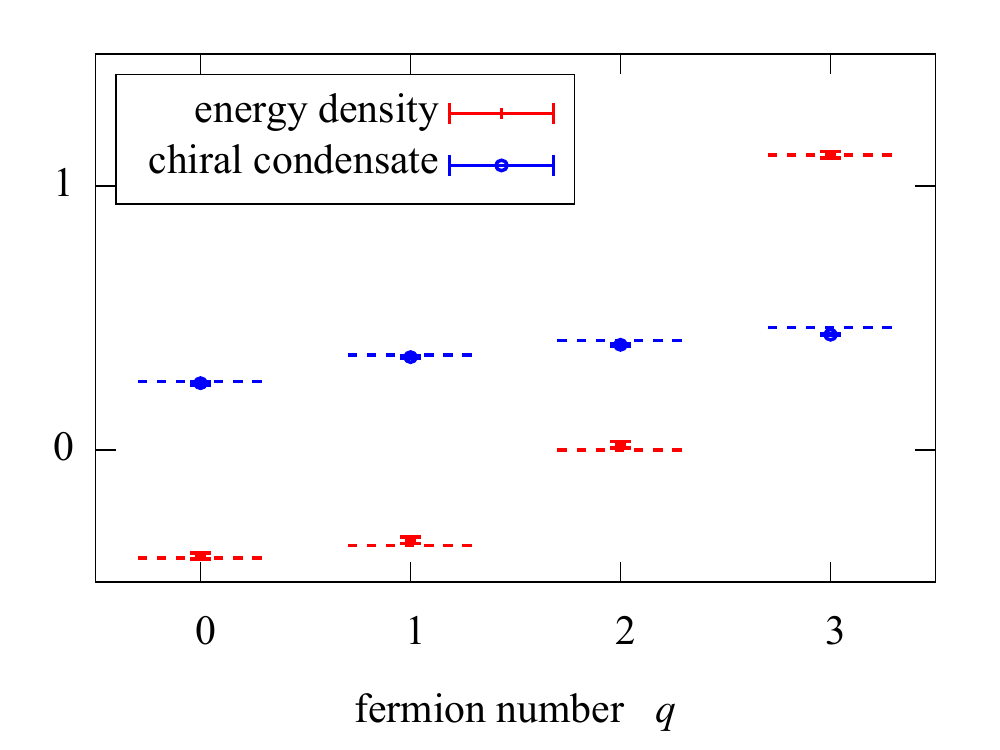}
\caption{\label{figQ}
The energy density (red) and the chiral condensate (blue).
The solid lines are obtained by the quantum variational calculation.
The dotted lines are the exact values obtained by the matrix diagonalization.
} 
\end{figure}

Once we obtain all the ground states $|\Psi(q)\rangle$, we can calculate physical observables as a function of particle number $q$.
For example, the energy density $E/N$ and the chiral condensate
\begin{equation}
 C = \frac{1}{N}\sum_{n=1}^{N} \langle\Psi(q)| (-1)^n \chi^\dagger_n \chi_n |\Psi(q)\rangle + \frac12
\end{equation} 
are shown in Fig.~\ref{figQ}.
The exact values are well reproduced.
Figure~\ref{figQ} is something like the famous plot in dense QCD.
In the study of dense QCD, we usually plot the chiral condensate as a function of chemical potential.
The horizontal axis is now replaced by particle number.

\section{Summary}

In this work, a computational strategy is designed for quantum lattice simulation at nonzero density.
Unfortunately, it is not applicable to the lattice QCD simulation in the current quantum computer.
The computational resource of the current quantum computer is quite limited.
It is not enough for QCD.
The strategy itself is however very general.
I hope that it becomes applicable to QCD someday in the future.

\section*{Acknowledgments}

The author was supported by JSPS KAKENHI Grant No.~19K03841.   
The author acknowledges the use of IBM Quantum services for this work.
The views expressed are those of the author, and do not reflect the official policy or position of IBM or the IBM Quantum team.

\end{document}